\documentstyle[12pt,aaspp4]{article}

\def\parcsec{{\tt ''}\mskip -7.6mu\,}   
\def\parcmin{{\tt '}\mskip -6.0mu.\,}   

\def\etal{{et~al.\ }}

\def\eg{{e.g.\ }}

\def\simlt{\ {\raise-.5ex\hbox{$\buildrel<\over\sim$}}\ }
\def\simgt{\ {\raise-.5ex\hbox{$\buildrel>\over\sim$}}\ }
\def\m6{Mrk~6}
\def\ROSAT{{\it ROSAT\/\ }}
\def\ASCA{{\it ASCA\/\ }}
\def\RXTE{{\it RXTE\/\ }}
\def\AXAF{{\it AXAF\/\ }}

\def\Astro-E{{\it Astro-E\/\ }}
\def\NIIf{{\rm N}\kern 0.1em{\sc ii}}
\def\OIIf{{\rm O}\kern 0.1em{\sc ii}}
\def\OIIIf{{\rm O}\kern 0.1em{\sc iii}}
\def\Feten{{\rm Fe}\kern 0.1em{\sc x}}
\def\Fe7{{\rm Fe}\kern 0.1em{\sc vii}}

\begin{document} 

\title{Heavy and Complex X-ray Absorption Towards the Nucleus of Markarian 6}

\author{J. J. Feldmeier and W. N. Brandt}
\affil{Department of Astronomy and Astrophysics, Penn State University,
525 Davey Lab, University Park, PA 16802}

\author {M. Elvis}
\affil{Harvard-Smithsonian Center for Astrophysics, 60 Garden Street, Cambridge, MA 02138}

\author{A. C. Fabian and K. Iwasawa}
\affil{Institute of Astronomy, Madingley Road, Cambridge CB3 0HA U.K.}

\and 

\author{S. Mathur}
\affil{Harvard-Smithsonian Center for Astrophysics, 60 Garden Street, Cambridge, MA 02138}

\begin{abstract}

We have used the \ASCA satellite to make the first X-ray spectra 
of Markarian~6, a bright Seyfert~1.5 galaxy with complex and variable 
permitted lines, an ionization cone, and remarkable radio structures. 
Our 0.6--9.5~keV spectra penetrate to the black hole core of this
Seyfert and reveal heavy and complex intrinsic X-ray absorption. Both 
total covering and single partial covering models fail to 
acceptably fit the observed absorption, and double partial covering or partial 
covering plus warm absorption appears to be required. The double partial 
covering model provides the best statistical fit to the data, and we 
measure large column densities of $\approx$ (3--20)$\times 10^{22}$~cm$^{-2}$
irrespective of the particular spectral model under consideration. 
These X-ray columns are over an order of magnitude larger than expected 
based on observations at longer wavelengths. Our data suggest that most of
the X-ray absorption occurs either in gas that has a relatively small amount 
of dust or in gas that is located within the Broad Line Region. The X-ray 
absorber may well be the putative `atmosphere' above the torus that 
collimates the ionization cone. 
We also detect an apparently broad 6.4~keV iron~K$\alpha$ line, and 
we present optical spectra demonstrating that the optical emission 
lines were in a representative state during our \ASCA observation. 

\end{abstract}


\section{Introduction}

X-ray observations of Seyfert galaxies penetrate into the centermost
regions of these vigorous objects and are highly effective at 
studying gas and dust along the line of sight to the black hole 
core. The nuclear X-ray source is generally thought to be even more 
compact than the Broad Line Region (BLR) and hence provides a
geometrically simple (when viewed from afar) emission region right 
at the heart of the Seyfert. X-rays are photoelectrically 
absorbed by neutral atomic gas 
and molecular gas in essentially the same manner, and most 
dust grains absorb X-rays as would an equivalent amount of 
neutral gas. Highly ionized gas can also be effectively probed
by means of oxygen and other X-ray absorption edges. 
In this paper, we use 0.6--9.5~keV spectra provided by the
Japan/USA {\it Advanced Satellite for Cosmology and Astrophysics}
({\it ASCA\/}; see Tanaka, Inoue \& Holt 1994) to 
perform the first detailed study of the  
X-ray emission and absorption properties of 
Markarian~6 (IC~450; 
$V \approx 14.2$, see Weedman 1972; 
$V_0=5784\pm 10$~km~s$^{-1}$,
see Meaburn, Whitehead \& Pedlar 1989). 
We also present results from nearly simultaneous optical spectroscopy. 
Mrk~6 is an S0 galaxy that contains a Seyfert 
nucleus of type~1.5 (e.g. Osterbrock \& Koski 1976 and references 
therein). Its optical hydrogen lines have a great deal of 
structure and have shown substantial profile variations
(e.g. Khachikian \& Weedman 1971, also see Adams 1972;
Rosenblatt et~al. 1992; Eracleous \& Halpern 1993), suggesting 
that much of the line emitting gas has a spatially coherent bulk 
velocity field. 

Mrk~6 is one of only three type~1 Seyferts which shows evidence for 
an `ionization cone' (Meaburn, Whitehead \& Pedlar 1989; 
Kukula \etal 1996; see Wilson \& Tsvetanov 1994 for a review of 
ionization cones). In the context of simple Seyfert unification 
models, the presence of an ionization cone in a type~1 Seyfert
is somewhat surprising. Seyfert~1s are thought to correspond 
to the geometry where we view the nucleus directly, and hence 
our line of sight should lie within the cone itself (we would
then see a halo morphology rather than a cone-like one). A 
possible explanation for the presence of ionization cones 
in some Seyfert~1s postulates that the obscuring torus 
has a lower density `atmosphere' (e.g. Evans \etal 1993; 
Kriss, Tsvetanov \& Davidsen 1994; Wilson 1996). Our line of
sight skims the surface of the torus and thus passes through
the atmosphere. The torus atmosphere is envisaged to have a column 
density of neutral/low-ionization material of 
$\approx 10^{20}$--$10^{21}$~cm$^{-2}$. Such a column 
is sufficient to be optically thick at all wavelengths 
between the Lyman edge and soft X-rays (the frequency range 
of the ionizing radiation), but it allows optical emission from
the BLR to pass through with only moderate
extinction. Thus we still see a type~1 optical spectrum
despite the fact that our line of sight lies outside the
ionization cone. 
\ASCA observations of the two other type~1 Seyferts with ionization 
cones, NGC~4151 and Mrk~766 (Weaver \etal 1994; Yaqoob \etal 1995; 
Leighly \etal 1996), show that they both have
complex low-energy X-ray absorption 
which could arise in torus atmospheres. 
Like Mrk~6, both NGC~4151 and Mrk~766 are Seyfert~1.5s
(see Osterbrock \& Koski 1976 and Osterbrock \& Martel 1993),
and all three objects may have their AGN aligned in a
fairly edge-on manner (e.g. section~3.2 of 
Evans et~al. 1993). A fairly edge-on orientation of
the AGN in Mrk~6 is further supported by the length and 
structure of its impressive radio jet (see Kukula et~al. 1996).
The large angular size of the Mrk~6 jet when compared to jets 
seen in other similar type~1 Seyferts suggests that it does not 
suffer from significant foreshortening by projection effects
(consistent with an edge-on orientation). 

Absorption in the nucleus of Mrk~6 has been studied at
radio, near-infrared and optical wavelengths. 
In the radio band 
(see Gallimore et~al. 1998 and references therein), 
H~{\sc i} (21~cm) absorption is detected only towards a bright, compact
radio feature located, in projection, $\approx 380$~pc 
($1^{\prime\prime}$) north of the optical nucleus (the column
towards this feature is $\sim 2.6\times 10^{21}$~cm$^{-2}$). 
The radio components nearest to the
AGN core did not show H~{\sc i} absorption with 
3$\sigma$ column density limits of  
$N_{\rm H}\simlt 4\times 10^{20}$~cm$^{-2}$ ($T_{\rm spin}$ / 100 K) ($\Delta v$ / 30 km s$^{-1}$), 
where $T_{\rm spin}$ is the spin temperature of the ground state
and $\Delta v$ is the FWHM of the absorption line. 
However, Gallimore et~al. (1998) 
did not detect a clear radio candidate for the AGN core, and thus 
it is not clear that radio measurements have constrained absorption 
along the line of sight to the AGN proper. 
Several near-infrared and optical observations have been 
performed in attempts to constrain the reddening to the 
Narrow-Line Region (NLR) and BLR of Mrk~6. 
Based on [O~{\sc ii}] and [S~{\sc ii}] line ratios, 
Malkan \& Oke (1983) suggested a NLR $E(B-V)$ of
$0.18\pm 0.08$ [here we have removed the $E(B-V)$ due to
extinction in our Galaxy]. Assuming a `Galactic' dust-to-gas 
ratio (Burstein \& Heiles 1978), this 
$E(B-V)$ corresponds to an absorption column density 
of $\approx (1.0\pm 0.4)\times 10^{21}$~cm$^{-2}$. 
McAlary \etal (1986) describe attempts to 
determine the BLR $E(B-V)$ but find that many of the
standard hydrogen lines appear to suffer from the effects
of high-density radiative recombination (which renders their 
ratios unreliable as reddening indicators). 
Rix \etal (1990) have performed Pa$\beta$ spectroscopy for
Mrk~6 and argue that use of this line does allow reliable
reddening constraints to be placed upon the BLR. They find 
that the extinction towards the BLR is no more, and is quite 
probably less, than $A_V=2$ [corresponding to 
$E(B-V)=0.7$ and an absorption column density of 
$\approx 4\times 10^{21}$~cm$^{-2}$].  
Intrinsic extinction and optical polarization are often found
together in Seyfert galaxies. Berriman (1989) presents
polarimetric measurements for \m6 and finds that 
it does appear to have some intrinsic polarization
(at a level of $0.52\pm 0.15$\%). However, this level
of polarization is not outstanding for type~1 Seyfert galaxies
(compare with the 67 other Seyfert~1s in Berriman 1989). 

Despite its proximity, brightness and fascinating 
properties at a variety of wavelengths, \m6 
has been very poorly studied in the X-ray regime. 
To our knowledge, the only X-ray detections of \m6 are
a possible {\it Uhuru\/} detection (as 4U~0638+74; Forman et~al. 1978),
a probable {\it HEAO-1\/} detection (as 1H~0641+741; Grossan 1992), 
and a secure \ROSAT All-Sky Survey (RASS) detection 
(as 1RXS~J065209.8+742537; Voges et~al. 1996).  
Although the {\it HEAO-1\/} 5~keV flux
density ($1.32\pm 0.13$~$\mu$Jy) of \m6 was 79\% that of
the bright and well-known Seyfert~1 galaxy
NGC~3783 (see chapter~2 of Grossan 1992), 
its flux in the \ROSAT band was low, with a RASS count rate 
of only $0.062\pm 0.012$~count~s$^{-1}$ (for comparison, the
RASS count rate of NGC~3783 was $1.4\pm 0.1$~count~s$^{-1}$). 
This discrepancy could be explained if the \ROSAT flux were 
reduced due to heavy intrinsic absorption of soft X-rays in Mrk~6. 

We proposed and performed an \ASCA observation of \m6 with 
several goals in mind. First of all, we wanted to define its basic 
X-ray properties (e.g. 0.6--9.5 keV flux) far better than has been
possible to date. In addition, we wanted to quantify the 
amount and nature of any intrinsic X-ray absorption so that it
could be compared with absorption at other wavelengths.
We also wanted to search for flux/spectral variability
and study iron K$\alpha$ line emission. Detection and fitting of a 
relativistic iron K$\alpha$ line might, for example, allow the
inclination of the inner accretion disk to be constrained. 
This is of particular interest for Mrk~6 since we have independent 
indications that the AGN is viewed in a fairly edge-on manner 
(see above).  We obtained nearly simultaneous optical spectra
to constrain the state of the optical lines during our \ASCA
observation.

We adopt a distance to \m6 of 77 Mpc 
($H_{0} = $ $75~$km~s$^{-1}~$Mpc$^{-1}$).  
The Galactic neutral hydrogen column 
density towards \m6 is 
$6.4\times 10^{20}$~cm$^{-2}$ 
(Stark et~al. 1992) and has an expected error of $\simlt 1 \times
10^{20}~$cm$^{2}$ (cf. Elvis, Lockman \& Fassnacht 1994).

\section{X-ray Observations and Data Analysis}

\subsection{X-ray Observation Details and Data Reduction}

\m6 was observed with \ASCA on 7--8 April 1997  
during the AO-5 observing round. 
Both Solid-state Imaging Spectrometer CCD detectors (SIS0 and SIS1) 
and both Gas Imaging Spectrometer scintillation proportional 
counters (GIS2 and GIS3) were operated.  The SIS detectors were
operated in 1 CCD mode, and the best calibrated SIS chips were used
(chip 1 for SIS0 and chip 3 for SIS1).  The SIS chip temperatures
varied from $-60.1$ to $-61.6$ degrees Celsius 
for the SIS0 detector, and from $-59.4$ to $-61.4$ degrees Celsius 
for the SIS1 detector.  The lower level discriminator was set to 0.47~keV 
for the SIS detectors.  The GIS were operated in PH mode.  

We have used the `Revision~2' processed data from Goddard Space Flight Center
(GSFC) for the X-ray analysis below (see Pier 1997 for a description
of Revision~2 processing), and data reduction was performed using
{\sc ftools} and 
{\sc xselect}. We have used the charge transfer inefficiency (CTI)
table released on 11~March~1997 [{\tt sisph2pi\_110397.fits};
see Dotani et~al. 1995 and section~7.7.1 of the
AO-6 \ASCA Technical Description (AN 97-OSS-02)
for discussions of CTI].  We have adopted the  
GSFC Revision 2 screening criteria.  After data screening, 
the total exposure times are the following: 
36.7 ks for SIS0, 36.6 ks for SIS1, 36.0 ks for GIS2 and 36.6 ks for 
GIS3.  We have extracted source photons using 
a circular region of radius $3\parcmin9$ for SIS0, 
an elliptical region with semimajor axis $4\parcmin4$ 
and semiminor axis $3\parcmin3$ for SIS1,
and circular regions with radii of $5\parcmin0$ for the GIS.  
The elliptical region for SIS1 provides the best
possible ratio of source to background photons, and it
prevents the extraction region from extending past the 
edge of the CCD chip. 

\subsection{Variability Analysis}

To search for flux variations, we binned the data from each
detector into 128 s bins, requiring all such bins to be fully
exposed.  For comparison to the \ASCA 
variability results on Seyfert~1 galaxies presented by 
Nandra \etal (1997a), we use identical energy bands for
our analysis: SIS0 + SIS1 full band (0.5--10 keV), 
SIS0 + SIS1 soft band (0.5--2 keV), SIS0 + SIS1 hard band 
(2--10 keV), and GIS2 + GIS3 hard band (2--10 keV).  
The mean count rates per second for each 
energy band are: 0.227 (SIS full), 0.049
(SIS soft), 0.178 (SIS hard) and 0.153 (GIS hard).  We tested 
for variability by means of a $\chi^{2}$ test against the hypothesis that
the count rate was constant.  The reduced $\chi^{2}$ values
found for each band are consistent with no variability,
and any sustained variability has an amplitude of less than
$\approx 20$\%. We also calculated the 
normalized `excess variance' ($\sigma^{2}_{\rm RMS}$) as 
defined by Nandra \etal (1997a).
The excess variance results, given in units of 
10$^{-2}~\sigma^{2}_{\rm RMS}$ (in order to be directly 
comparable to the results of Nandra \etal 1997a), are 
$0.46 \pm 0.37$ (SIS full), $-0.02 \pm 0.44$ (GIS hard), and $0.71 \pm 0.49$  
(SIS hard) where the error bars are at the 68\% confidence level.  
Since the count rate in the SIS soft band was very small, the excess 
variance results were unreliable, and therefore we do not 
report them.  The excess variances are compatible with minimal or no 
variability. 

\subsection{Spectral Analysis}

\subsubsection{Preparation of Spectra and Fitting Details} 

Since there is no strong evidence for variability, we have 
extracted 0.6--9.5~keV SIS and 1.0--9.5~keV GIS spectra 
using all of the acceptable exposure time.
We have grouped these spectra so that there are a minimum of 
15 photons per spectral data point, to allow the use of 
chi-squared fitting techniques.  We have used the {\sc sisrmg} 
software to generate our SIS redistribution matrix 
files (rmf), and we use the GIS rmf from 1995 March 6.
We generate our ancillary response files (arf) using
the {\sc ascaarf} software, and we perform spectral modeling using 
{\sc xspec} (Version 10; Arnaud 1996).  Unless 
stated otherwise, all errors are quoted at the 90.0\% 
confidence level for one parameter of interest 
($\Delta \chi^{2} = 2.71$).

We first performed spectral fitting for each of the \ASCA detectors
separately, and we obtained results that were consistent to within
the errors.  We have therefore jointly fitted the spectra from all
four detectors, and we detail these results below.  In such joint fitting, 
we allow the normalization for each detector to be free (to allow for
small calibration uncertainties in the absolute normalizations between
detectors; this is currently standard practice when performing \ASCA
analysis), but we tie together all other fit parameters across the four 
\ASCA detectors.  When we quote fluxes and equivalent widths 
below, we shall quote them for the SIS0 detector. 

We adopt `solar metallicity' and the `solar photosphere' abundance pattern 
from Table~2 of Anders \& Grevesse (1989) for our fitting of intrinsic 
absorption models (unless stated otherwise). 

\subsubsection{Basic Spectral Fitting} 

In this section, we shall number the various X-ray spectral models as they 
are introduced. 
For the main models discussed below, the relevant model parameters are given 
in Table~1. 
We compare several of the spectral models discussed below in Figure~1a. 
We also display the SIS0 residuals for Models~1--7 in Figure~2. These 
show the systematic trends in the residuals that are not apparent from 
only the numerical value of $\chi^2_\nu$.

We began by fitting the spectra with a power-law plus neutral Galactic 
absorption model (Model~1). $N_{\rm H}$ was constrained to lie within the 
radio-determined Galactic column density range of 
(5.4--7.4)$\times 10^{20}$~cm$^{-2}$. This model gives a reduced 
$\chi^{2}$ of 1.33 for 778 degrees of freedom, and hence it is rejected 
with over 99\% confidence according to the P($\chi^{2}$ $\mid$ $\nu$)
function as defined in section~6.2 of Press et~al. (1989). 
The model left systematic residuals throughout
the spectrum, and the fitted power law was extremely flat 
($\Gamma = 0.01^{+0.03}_{-0.03}$). Such a flat intrinsic power law is not 
expected from a Seyfert~1 galaxy. Compilations of intrinsic \ASCA photon 
indices for Seyfert~1 galaxies (e.g. Brandt, Mathur \& Elvis 1997; 
Nandra \etal 1997b) show almost none less than 1.6. The flat photon
index and systematic residuals could arise as the result of intrinsic
absorption that has not been taken into account. 
Therefore, we next added intrinsic absorption in \m6 itself to the power-law 
plus Galactic absorption model (Model~2).  Although this model 
did find a large amount of intrinsic absorption in \m6 
[$N_{\rm H,\m6}=(1.19^{+0.16}_{-0.15})\times 10^{22}$~cm$^{-2}$], 
it is still a poor fit to the data with a reduced $\chi^{2}$ 
value of 1.08. For 777 degrees of freedom, it is ruled out at the 
94.0\% confidence level. The model left obvious 
positive residuals (i.e. the data points were above the model)
at the soft end of the X-ray spectrum, and again
the best-fitting power law is flatter than expected for a 
Seyfert~1 galaxy ($\Gamma = 0.56^{+0.07}_{-0.09}$). 
If we constrain the power law in Model~2 to have a photon index
of at least 1.6 (Model~3), we obtain
$N_{\rm H,\m6}=(3.66^{+0.17}_{-0.16})\times 10^{22}$~cm$^{-2}$
and $\chi^2_\nu=1.54$ for 777 degrees of freedom. 
Large systematic residuals are present throughout the spectrum
(see Figure~3), and the fit is ruled out with over 99\% 
confidence.   

We next attempted to fit the spectra with models that have been 
successful in studies of other similar Seyfert galaxies.      
In the Seyfert 1.5 galaxy NGC~4151, a potentially similar object
(see \S1), a successful model is that of leakage through a 
partially covering absorber (\eg Weaver \etal 1994). 
Partial covering is also found to be important more generally
in the X-ray spectra of intermediate-type Seyferts (e.g. Forster 1998). 
We therefore tried a model consisting of a power law, Galactic 
absorption, and a partially covering absorber at 
the redshift of \m6 (Model~4).  Here and hereafter, we constrain
the power-law photon index to be at least 1.6 (see the previous paragraph
for justification).  Model~4 substantially 
improved the fit to the data over Model~3 ($\Delta \chi^{2} = -357.0$), 
with a reduced $\chi^{2}$ of 1.09 for 776 degrees of freedom.  However,
it still left systematic residuals in the soft and hard portions of the 
spectrum (see Figure~4) and was ruled out at the 95.1\% confidence level.   

The partial covering in Model~4 is physically equivalent to two
components of X-ray emission: (1) a direct, but attenuated, component
(perhaps viewed through the torus atmosphere) and (2) a scattered, 
unattenuated component.  There is no reason {\it a priori\/} that 
the X-ray emission should follow only two light paths.  In reality, there 
could be multiple attenuated lines of sight due to, for example, scattering
followed by some absorption.  In order to better approximate these multiple 
paths, we fit the data with a model consisting of a power law, Galactic
absorption, and two partial covering absorption components (Model~5).  
The second partial covering component has a much
larger column density than the first [$(29.8^{+4.4}_{-9.9})\times 
10^{22}$~cm$^{-2}$ compared to 
$(3.33^{+0.29}_{-0.16})\times 10^{22}$~cm$^{-2}$], but a smaller
covering percentage ($55.7^{+5.5}_{-2.9}$\% compared to 
$92.7^{+1.1}_{-0.7}$\%).  Model~5 
provides a substantial improvement in fit quality relative to Model~4 
($\Delta\chi^{2} = -128.2$) with the addition of 
only two more parameters.  The model is now a good fit 
to the data (reduced $\chi^{2} = 0.923$), and no large-scale 
systematic residuals can be seen.

Spectral features due to ionized `warm' absorption have also been
investigated in the X-ray spectra of NGC~4151 and Mrk~766.
The strongest features associated with warm 
absorbers are K-edges from \ion{O}{7} (0.739~keV)
and \ion{O}{8} (0.871~keV), but additional edge and line features are
also thought to be present. 
We have fit the data with a self-consistent photoionized absorber/emitter 
model constructed from {\sc cloudy} (Ferland 1996) calculations 
(Model~6; see Reynolds \& Fabian 1995 for details of the photoionization
calculations). The two free parameters in this model are the
ionization parameter $\xi$ and the column density of ionized gas
$N_{\rm H,warm}$. We obtain 
$\xi=74.1^{+1.6}_{-1.2}$~erg~cm~s$^{-1}$ and
$\log (N_{\rm H,warm})=22.95^{+0.02}_{-0.02}$. 
The model gives $\chi^2_\nu=1.12$ for 776 degrees of freedom, and
hence it is rejected with $>99$\% confidence. 
It is also possible that the spectrum has a combination of partial
covering and warm absorption. To test this, we have fit 
a model consisting of the single partial covering of Model~3
plus the ionized absorption of Model~6 (Model~7; see Table~1 for details). 
This model provides a statistically acceptable fit to the data. 

Out of all the models presented so far, only Models~5 and 7 
provide statistically and physically acceptable fits to the data.  
These models have the same number of degrees of freedom, but Model~5 
provides $\Delta\chi^{2} = -40.7$ relative to Model~7. Although
we cannot rule out the partial covering plus warm absorber model, the 
double partial covering model is favored by the data, and we have 
adopted it for further analysis.  

We next searched for the presence of the iron~K$\alpha$ 
fluorescence line. This line has been seen in X-ray spectra of
many Seyfert~1 nuclei, and we observe systematic positive
residuals at the expected energy (see Figure~4). 
We added a narrow Gaussian line to Model~5 
with a rest-frame energy of 6.4 keV and a fixed width of $\sigma$ = 20 eV 
(Model~8).  This model provides an improvement in fit quality of 
$\Delta\chi^{2} = -10.5$ which is significant with over 99\% confidence
according to the $F$-test (see tables C-5 and C-6 of 
Bevington \& Robinson 1992). The iron~K$\alpha$ line for Model~8 has an
equivalent width of $155^{+92}_{-92}$~eV.  With the presence of the 
iron~K$\alpha$ line established, we remove our constraint on the 
line width and allow {\sc xspec} to find the best fitting width 
(Model~9).  The best-fitting intrinsic width for the Gaussian line 
is $235^{+327}_{-168}$ eV, suggesting that the line may have some breadth.  
We find a corresponding equivalent width of 
$251^{+518}_{-175}$~eV.  We have also 
experimented with models of a relativistic iron line from the inner part
of an accretion disk, but we find the parameters of such models (\eg 
orientation) to be poorly constrained due to the limited 
photon statistics for the line.

\subsubsection{Testing of Alternative Continuum Shapes} 

To examine the robustness of the above results, we have investigated 
the effects of including additional X-ray continuum components. 
Our goal is to examine whether a physically-reasonable model with a 
different X-ray continuum shape can significantly alter our main 
conclusions. 

We began by considering continuum emission from a 
soft X-ray excess that is made by diffuse 
hot gas located outside the intrinsic absorber
(e.g. Wilson et~al. 1992; Weaver et~al. 1995). Such gas might be
heated by starburst activity in the galactic disk 
(cf. Baum et~al. 1993). To model this gas, we have added a 
Raymond-Smith thermal plasma (RSTP) component to Models~3, 4, 6 and 7. 
We have restricted the temperature of the RSTP to 
be $<1$~keV based on X-ray observations of starburst activity in
other galaxies, and we allow the metallicity to vary freely.
Models 3 and 6 with the RSTP are statistically rejected with 
over 99\% confidence, and Model~4 with the RSTP is not significantly 
better than Model~4 without it (using the $F$-test). Model~7 with the
RSTP gives a metallicity of 0, and this is unphysical. If we fix
the metallicity at solar, we obtain $T\approx 0.9$~keV 
with $\chi^2=734.5$ for 772 degrees of freedom. This is a
statistically acceptable fit, although it is not as good as 
that for Model~5 ($\Delta\chi^2=20.2$, and Model~5 has 2 fewer
degrees of freedom). In addition, this model still leaves
weak systematic residuals above 1~keV. 

We then considered soft X-ray excess emission originating from the
immediate vicinity of the black hole. We parameterized this soft 
excess emission in three different ways: blackbody emission, 
bremsstrahlung emission and RSTP emission. Based on observations
of many other Seyfert~1 galaxies, we constrained the 
temperatures of the soft excess models so that they did not 
dominate the flux above 1.5~keV. For Model~3,
adding soft excess components failed to improve its fit to 
statistically acceptable levels. The other models tested either
did not improve the fit significantly according to the $F$-test
or were not as good as the Model~7+RSTP model of the previous
paragraph. 

We have also investigated the effects that the Compton reflection continuum
might have on our basic absorption results. We started by fitting
versions of Model~1 and Model~3 with the simple power law replaced
by an exponentially cut-off power law reflected from neutral 
material (the `pexrav' model in {\sc xspec}). We constrained the
cut-off energy to be $>100$~keV and the intrinsic photon index
to be in the range 1.6--2.2. Both of these models are statistically
rejected (the first at $>99$\% confidence and the second at 
$>97.5$\% confidence) and gave implausibly high reflection fractions
($>10$). We then fit a version of Model~4 with the simple power 
law replaced by the pexrav model. If we do not constrain the reflection
fraction, it becomes unphysically large ($\approx 9.4$). If we 
constrain the reflection fraction to be $<2$, we 
obtain $\chi^2=780.8$ for 774 degrees of
freedom. The fit is still not as good as that for Model~5 
($\Delta\chi^2=66.5$), and systematic residuals similar to those seen
in Figure~4 are still apparent. The reflection fraction `pegs' at 
the rather high value of 2, and the statistically best-fitting model 
is for face-on reflection.

To summarize, inclusion of a soft X-ray excess does not appear to be
able to alter our main results regarding the presence of heavy and
complex X-ray absorption. While a soft X-ray excess may be present,
none of the soft excess models investigated above improves the fit
to where it is as good as Model~5. A Compton reflection continuum
may also be present, but again it appears to be unable to alter our 
main absorption results. 

\subsubsection{Additional Safety and Robustness Checking} 

We have verified that reasonable deviations from `solar photosphere' 
metallicity and abundances do not change our basic absorption results. 
For example, we have fit a version of Model~3 where we allow both the 
global metallicity and the individual abundance ratios to vary between 
0.2--5 times the solar values (cf. Wheeler, Sneden \& Truran 1989). 
This model can be statistically rejected with $>99$\% confidence. 
Global changes in metallicity with the relative abundances
fixed at the solar pattern 
(from Table~2 of Anders \& Grevesse 1989) do not
have a large effect on $\chi^2_\nu$ for all models
discussed above. This is because our \ASCA spectra do not
tightly constrain low-energy absorption by H and He, and
thus to first order an increase in global metallicity
can be offset by a decrease in column density. As a result, 
the basic line of argumentation in \S 2.3.2 still applies,
and implausibly large metallicities would be needed to resolve
the X-ray versus near-infrared/optical absorption discrepancy
discussed in \S 4.2.1. 

To further scrutinize our results, we now focus
on the power-law photon indices of our models.  In all the models where 
the photon index was constrained to be 
greater than 1.6, the best-fitting parameters were always found when 
the index was `pegged' at the minimum value allowed.  As
has been stated previously, smaller photon indices are unlikely,
yet we would also expect that a physically correct model would not 
be driven past a hard limit in any reasonable parameter.  
To investigate this matter, we re-fit Model~5 allowing 
the photon index to be free.  The best-fitting model has a 
photon index of $\Gamma = 1.46^{+0.23}_{-0.32}$.  The best fitting
photon index is somewhat smaller than the range observed in Seyfert 1s 
($\Gamma \approx$ 1.6--2.2), but it is statistically consistent given
the errors.  To further examine the limits on the photon index, 
we re-fit Model~5 fixing the 
photon index at multiple values throughout the range 
observed for Seyfert~1s.  Although low values 
of $\Gamma$ are statistically favored, none of the models
could be statistically rejected; the 
worst fit (for $\Gamma = 2.2$) was rejected 
at only the 19.3\% confidence level.  The 
complex absorption can compensate for steeper intrinsic 
power laws, and hence the intrinsic photon index is 
poorly constrained.  We stress that steeper power laws
only {\it increase\/} the amount of absorption required, so the 
presence of heavy absorption is extremely robust. 

Finally, we have repeated our analysis several times
using different choices for the CTI table and extraction 
regions, and none of our main results materially depends upon 
the details of our specific choices. 

\subsubsection{Fluxes and Luminosities} 

We have used our best-fitting model, 
Model~9, to compute X-ray fluxes and 
luminosities for \m6.  We find an observed 2--10 keV 
X-ray flux of $1.0 \times 10^{-11}$ ergs cm$^{-2}~$s$^{-1}$ 
and an unabsorbed flux of $1.7 \times 
10^{-11}$ ergs cm~$^{-2}~$s$^{-1}$.  Due to the strong
absorption, the 0.5--2 keV flux found from the model is
considerably less: $3.8 \times 10^{-13}$ ergs cm$^{-2}~$s$^{-1}$
for the absorbed flux and $7.7 \times 10^{-12}$ ergs cm$^{-2}~$s$^{-1}$
for the unabsorbed flux.  Assuming the distance to \m6 stated
in \S 1, the unabsorbed X-ray luminosity is 
$1.2 \times 10^{43}$~ergs~s$^{-1}$ in the 2--10 keV band and 
$5.4 \times 10^{42}$~ergs~s$^{-1}$ in the 0.5--2 keV band.   

\section{Optical Observations and Data Analysis}
\subsection{Optical Observation Details and Data Reduction}

Optical spectra were obtained 33~days before (5~March~1997)
and 1~day after (9~April~1997) 
the \ASCA observation (see Table~2 for observation details). 
The spectra were bias-subtracted, flat-fielded and extracted 
from the CCD frames using {\sc iraf} (the {\sc apall} task was
used for the extraction). The spectra were then 
wavelength calibrated using He-Ar and 
Ne lamps for the FAST data 
and a He-Ne-Ar-Hg-Cd lamp for the MMT data.
Finally, the spectra were flux calibrated. 
The spectrophotometric standard HZ~44 was used to calibrate the
FAST observation, and the spectrophotometric standards
Hiltner~600 and PG~0823+546 were used for the MMT observation
(see Massey et~al. 1988 for a discussion of these photometric
standards). However, due to poor seeing and slightly 
non-photometric conditions, we primarily use 
these calibrations to determine relative fluxes rather than 
absolute ones.  The final reduced spectra for \m6 are shown in 
Figure~5.    

\subsection{Optical Analysis and Comparison with Historical Data}

Our primary intention in this optical analysis is not to measure 
flux ratios for all the spectral lines of \m6 (see Koski 1978 and 
Malkan \& Oke 1983 for detailed spectrophotometry), but instead 
to merely ensure that the variable optical spectrum was in 
a representative (rather than unusual) state during 
the \ASCA observation. This is important for establishing the
general applicability of our \ASCA results; an atypical
optical spectrum might suggest that we had observed \m6 during
a time when its internal X-ray absorption 
was unusually high or low. Our second optical observation was only 
one day after the \ASCA observation, and there is no historical 
evidence for strong optical variability on such a short timescale
(e.g. Eracleous \& Halpern 1993).

To characterize the optical state of \m6, we first used the 
flux ratio of the broad to narrow components of the H${\beta}$ spectral 
line.  This ratio has the advantages of reflecting the nuclear 
activity of \m6, being well observed historically, and being 
relatively insensitive to flux calibration uncertainties.  
Historically, the broad to narrow line ratio has varied from a 
maximum of 4.5 (see tables~2 and 13 of 
Rosenblatt \etal 1992) to a minimum of 
1.2 (see section~III of McAlary \etal 1986).  
To obtain the narrow line flux for our spectra, 
we de-blended it from the broad component, assuming that the narrow 
line component was fit with a Gaussian profile of fixed width found
from the unresolved [\OIIf]~$\lambda$~3727 spectral line.  
To find the broad line flux, 
we integrated over the entire H$\beta$ profile, after 
subtracting the narrow line component and the 
[\OIIIf]~$\lambda$~4959 line, which is partially blended 
with the broad component of H$\beta$.  We measure  
$\frac{F(H\beta_{\rm broad})}{F(H\beta_{\rm narrow})}$ 
to be $3.3 \pm 0.5$ for the 
5~March~1997 observation and $3.5 \pm 0.9$ for the 9~April~1997 observation.  
The error bars are the 1$\sigma$ errors from setting the continuum level.
The measured ratios are consistent with the  
historical range of activity and support the idea 
that \m6 was in a reasonably normal optical 
state at the time of the \ASCA observation.  

We have also determined approximate values for the Balmer decrement 
for the sum of the broad and narrow line components, after first 
deblending the [\NIIf] lines from the H$\alpha$ emission.  We find 
$\frac{F(H\alpha)}{F(H\beta)} = 6.1\pm 1.2$ for the 5~March~1997 observation and
$\frac{F(H\alpha)}{F(H\beta)} = 5.8\pm 1.4$ for the 9~April~1997 observation, 
and the errors on these
measurements are dominated by the uncertainties involved in flux 
calibration. Koski (1978) found 
$\frac{F(H\alpha)}{F(H\beta)} \approx 7.3$ 
and Malkan \& Oke (1983) found 
$\frac{F(H\alpha)}{F(H\beta)} \approx 7.2$, 
so again our spectra appear to be reasonably consistent
with the historical range of activity for \m6.

Finally, we convolved our spectra with a Johnson $V$ band filter response
curve (Bessell 1990) to find an approximate $V$ magnitude. We find 
$V=14.6\pm 0.5$ for the 5~March~1997 observation and 
$V=14.9\pm 0.5$ for the 9~April~1997 observation, where the 
large error bars are due to the small size of our spectrograph
aperture and the non-photometric conditions. These magnitudes
appear to be consistent with the historical range of activity. 
For comparison, Weedman (1972) found $V=14.2$. 


\section{Discussion and Conclusions}

The most significant result of this work is the detection of
heavy and complex X-ray absorption towards the nucleus of \m6. 
All of our acceptable models clearly show evidence for this
absorption, and its presence appears to be extremely robust. 
Below we will discuss the interpretation of this X-ray absorption, and 
we will compare it to absorption seen at other wavelengths. 


\subsection{Comparisons of X-ray, [O~{\sc iii}]~$\lambda$~5007 and Far-Infrared Fluxes}

Has our \ASCA observation penetrated all the way to the black hole
region of \m6, or is this region still blocked by even thicker 
X-ray absorption that \ASCA cannot penetrate? 
We do not observe X-ray flux variations, so we are not 
able to use the presence of rapid X-ray variability as an argument 
that we are directly observing the black hole region. 
However, we can compare the measured X-ray flux to fluxes at other 
wavelengths to address this issue. For example, Mulchaey \etal (1994) 
have studied a sample of Seyfert galaxies and discuss fairly strong 
correlations between the 2--10~keV flux (absorption corrected), 
the [\OIIIf]~$\lambda$~5007 
flux, and the far-infrared flux. Using these correlations, the expected 
2--10~keV flux can be predicted from the [\OIIIf]~$\lambda$~5007 and
far-infrared fluxes. Since it is believed that the 
[\OIIIf]~$\lambda$~5007 and 
far-infrared fluxes are reasonably isotropic properties of Seyferts, 
a large deficit of observed 2--10~keV flux compared to that
predicted might indicate the presence of additional extremely thick
absorption that \ASCA cannot penetrate. In this case, we might be
observing X-ray flux that has been scattered around the extremely
thick absorption and only suffers somewhat lighter absorption. 

Using the [\OIIIf]~$\lambda$~5007 and far-infrared fluxes given 
for \m6 in Mulchaey \etal (1994) and applying 
the correlations found therein, we predict 2--10~keV 
fluxes of $\approx 3\times 10^{-11}$~erg~cm$^{-2}$~s$^{-1}$ 
from the [\OIIIf]~$\lambda$~5007 flux and 
$\approx 2\times 10^{-11}$~erg~cm$^{-2}$~s$^{-1}$ from the
far-infrared flux.  
Our absorption-corrected 2--10~keV flux from \ASCA of
$1.7\times 10^{-11}$~erg~cm$^{-2}$~s$^{-1}$ 
is in reasonably good agreement with these values. 
Given the intrinsic scatter of the Mulchaey \etal (1994) correlations 
and potential variability of the X-ray flux, we find no significant
discrepancy between the predicted and observed 2--10~keV fluxes. 
In addition, we note that Mrk~6 has a complex [\OIIIf]~$\lambda$~5007 
morphology due to its ionization cone (e.g. Meaburn, Whitehead \& Pedlar 1989), 
and thus the measured [\OIIIf]~$\lambda$~5007 flux from an optical spectrum 
can depend upon the position angle of the slit on the sky. We see this effect 
in our optical spectra at the $\sim 40$\% level, and it may 
at least partially explain the [\OIIIf]~$\lambda$~5007 discrepancies
noted by McAlary et~al. (1986). In summary, the observed X-ray flux is 
entirely consistent with the idea that we have penetrated all the way to 
the black hole core of \m6. The probable broad iron~K$\alpha$ line discussed 
in \S 2 is further suggestive evidence that we are directly observing 
the black hole region. 


\subsection{The Physical Properties and Location of the Absorber}

\subsubsection{The X-ray Versus Near-Infrared/Optical Absorption Discrepancy}

Independent of the details of the spectral model under consideration, 
the X-ray absorption we observe appears to be about an order 
of magnitude larger than that expected based on earlier studies at 
longer wavelengths. Our best model (Model~9) involves double 
partial covering with column densities of
$\approx 3\times 10^{22}$~cm$^{-2}$ and
$\approx 2\times 10^{23}$~cm$^{-2}$. 
Rix et~al. (1990) have argued using 
near-infrared/optical data that the extinction to the BLR is no 
more, and is quite probably less, than 
$A_V=2$ (corresponding to a column density of 
$\approx 4\times 10^{21}$~cm$^{-2}$ for a `Galactic' 
dust-to-gas ratio; see \S 1 for details). 
The discrepancy between X-ray absorption and 
near-infrared/optical extinction
cannot be removed by appealing to the effects that 
high-density radiative recombination can have on hydrogen lines; 
correcting for such effects would only decrease the inferred 
near-infrared/optical extinction and thereby exacerbate the 
discrepancy. Similar absorption discrepancies have been 
observed in other Seyferts
(e.g. section~IV of Maccacaro, Perola \& Elvis 1982; 
section~7.2 of Turner \& Pounds 1989;
section~4.1 Veilleux, Goodrich \& Hill 1997).

The X-ray versus near-infrared/optical absorption discrepancy 
for \m6 reemphasizes the importance of X-ray column density 
measurements for studies of absorption 
in Seyfert galaxies. The best efforts at
near-infrared and optical wavelengths appear to have failed to 
expose {\it most\/} of the material along the line of sight to the 
black hole in \m6. To resolve the absorption discrepancy, one is 
forced to conclude that 
(1) most of the absorbing material lies within the BLR where it
could not be detected by earlier observations
and/or
(2) the X-ray absorption arises in relatively dust-free gas which causes
X-ray photoelectric absorption but not optical extinction
(see Maccacaro et~al. 1982 for further discussion). 
Note that these two possibilities need not be mutually exclusive;
gas lying within the BLR is quite likely to be relatively dust-free.
In fact, the most natural location for a large amount of relatively  
dust-free gas would be within the dust sublimation radius of the 
central engine (or perhaps somewhat outside of this region if
photoionization-driven sputtering is relevant; see section~2
of Krolik 1996). Following section~5.3 of Laor \& Draine (1993), the 
dust sublimation radius should be $\sim 25$ light days for \m6.  

We also comment that even the smallest X-ray column density is
$\simgt 75$ times larger than that for the radio components 
nearest to the AGN core (see \S 1 and Gallimore et~al. 1998). Our
interpretation here is that the currently available radio
data (the best of which have angular resolutions that 
correspond to $\sim 60$~pc) simply do not probe the absorption 
to the centermost regions. Note that there is currently no radio 
candidate for the central optical continuum source. 


\subsubsection{The Location of the X-ray Absorber}

In the context of the simplest and most popular version of the 
unified model for Seyfert galaxies, the heavy X-ray absorption we 
observe from \m6 might be associated with a compact ($<10$~pc), 
obscuring torus. However, before we interpret our data in terms 
of this model, we first critically examine other possible 
locations for absorber. 

Malkan, Gorjian, \& Tam (1998; hereafter MGT98) have proposed an 
alternative to the standard unified scheme for Seyfert galaxies known 
as the Galactic Dust Model (GDM). In the GDM, most of the absorption 
seen in Seyfert galaxies does not arise within a compact region but is
rather due to large-scale ($>100$~pc) dust lanes that have little
or no physical connection with the central engine.
Dust lanes are observed in {\it Hubble Space Telescope\/} images 
of some Seyfert galaxies, and MGT98 find an `irregular dust'
distribution for \m6. However, our data suggest that the GDM
is probably not applicable for the bulk of the observed X-ray
absorption in \m6. The evidence for X-ray absorption by relatively
dust-free gas suggests that the absorber is probably close to the 
central engine where dust can be sublimated and/or sputtered
(see above). In addition, the absorption
we see is of the partial covering (rather than the total covering) 
type, and we have suggested that this is due to the combination
of a direct, obscured line of sight and a scattered, unobscured line 
of sight. It would be somewhat of a geometrical challenge to 
arrange partial covering absorption by a large-scale dust lane
located hundreds of parsecs from the X-ray source. 
One would need both a fortuitous alignment of small-scale structure
in a dust lane as well as an unusually small dust-to-gas ratio
(or perhaps unusual dust that only causes grey extinction) for the
GDM to be applicable in this particular case. 


\subsubsection{X-ray Absorption in the Torus Atmosphere?}

As discussed in \S 1, the presence of an ionization cone in \m6 
has been used to suggest that we are viewing the nuclear region 
through a torus `atmosphere' that blocks most ionizing radiation but 
lets optical radiation pass through with only moderate extinction. 
The X-ray absorption we detect appears to be in general agreement
with this picture (see Figure~6), and the lack of high optical
polarization (see \S 1) could also result at least partially from 
this differential absorption as a function of wavelength (in addition, 
we note that there is probably significant dilution of any polarized 
optical light by emission from the host galaxy; see Rosenblatt et~al. 1992). 
However, the X-ray column densities we measure are substantially larger 
than the $\approx 10^{20}$--$10^{21}$~cm$^{-2}$ that have been typically 
associated with torus atmospheres (see section~7 of Wilson 1996).
It appears that we are either detecting an additional absorber located 
inside the torus atmosphere or that the torus atmosphere is composed
of relatively dust-free gas. In the second case, such gas could arise 
if dust were sublimated in the process of being evaporated off the torus. 

The two other Seyfert~1s with ionization cones, NGC~4151 and Mrk~766,
also show complex X-ray absorption. Of these two objects, 
\m6 more closely resembles NGC~4151 in its overall absorption 
properties. Mrk~766 appears to have a significantly lower neutral 
column density and is a bright soft X-ray source
(see Leighly \etal 1996 and references therein). \m6 and NGC~4151 
may be viewed at somewhat larger 
inclination angles so that our line of sight intercepts
thicker and less highly ionized material. Since both NGC~4151 
and Mrk~766 show strong X-ray spectral variability, we suspect that
\m6 may also show such variability. Systematic monitoring of 
X-ray spectral variability should give further clues about
processes in the torus atmosphere of \m6.  

\subsubsection {Implications for Other Wavelengths}

If the X-ray absorption is indeed due to a torus 
atmosphere, there are a number of implications for 
other observations of \m6.  Given the X-ray column densities 
we measure, it is likely that significant free-free absorption will be 
present at centimeter wavelengths. Applying the 
relations of Maloney (1996) for the free-free opacity of
a torus, we find that the optical depth at 21 cm 
due to free-free absorption is plausibly of order 
$\tau_{\rm ff}\sim$~5--100.  This large opacity could
naturally explain the non-detection of 
radio continuum emission from the nucleus by 
Gallimore \etal (1998) and the corresponding non-detection of 
H~{\sc i} absorption.

The torus atmosphere may also emit coronal lines in the near-infrared
and optical, and coronal lines (e.g. [\Fe7] $\lambda$ 6087) 
have already been observed from \m6 (Koski 1978). Pier \& Voit (1995) 
hypothesized that much of the coronal line emission from Seyferts 
comes from a layer just above the surface of the molecular torus,
although not all of the coronal line emission need come from this
region (see Murayama \& Taniguchi 1998). High-resolution spectroscopic 
observations of optical coronal lines such as [\Fe7] $\lambda$ 6087 
and [\Feten] $\lambda$ 6375 may allow a probe of the dynamics of
the torus atmosphere. Near-infrared coronal lines could also provide
constraints on its dust content.  Ferguson, 
Korista \& Ferland (1997) have calculated 
the strengths of near-infrared coronal lines over a large range 
of physical parameters relevant to Seyferts, and they argue that
these lines originate in a region of nearly 
dust-free gas. If large amounts of dust were present in the 
coronal line region, the near-infrared coronal lines of 
calcium and iron would be much weaker than generally 
observed due to shielding of the ionizing radiation and depletion 
onto dust.  Detection of these near-infrared coronal lines from \m6 could
thus provide additional evidence for a relatively dust-free torus
atmosphere. The near-infrared observations of \m6 to date have either 
had insufficient resolution, signal-to-noise, or wavelength coverage 
to search for these coronal lines, but a better study should be
possible with current near-infrared detectors.

Finally, if the torus atmosphere is as hot as $\sim 10^5$~K, thermal 
soft X-ray emission with a luminosity of $\sim 10^{41}$~erg~s$^{-1}$ 
is expected. Our current observations do not place any strong
constraints on this emission, and we expect that it will be difficult 
to separate any such emission from that of a circumnuclear starburst.


\subsection{Future X-ray Observations}
Based on our results, we identify several future X-ray 
observations that show particular promise for improving
our understanding of \m6.  First of all, X-ray spectra at 
higher energies (e.g. from \RXTE or {\it SAX})
would be helpful for measuring the intrinsic continuum shape. 
Our \ASCA spectra cannot tightly constrain the slope of the underlying
power law due to the heavy and complex absorption, and knowledge of
the intrinsic continuum slope would allow even tighter constraints 
to be placed on the X-ray absorption. Given the variability 
of the optical line emission and the fact that our line of sight
appears to intersect a torus atmosphere, searches for variability 
of the X-ray absorption properties are also
important to perform. Coordinated variability of the X-ray absorption
and optical line profiles might be especially revealing. 
The {\it XMM\/}, \Astro-E and {\it Constellation-X\/} 
missions will allow high-quality
spectroscopy of the iron~K$\alpha$ line emission, and it is important
to constrain the inner accretion disk inclination by fitting
the profile of the apparently broad line (see \S1 for
further discussion). Finally, \AXAF imaging would be useful to
constrain and/or study any extended soft X-ray emission 
associated with starburst activity. 

  
\section{Acknowledgments}

We thank P. Berlind, A. Dobryzcki, J. Huchra and J. Mader 
for their effort in obtaining the \m6 optical spectra.  
We thank K. Arnaud, M. Eracleous, K. Forster, K. Mukai and
D. Weedman for helpful discussions. 
We also thank the referee, J. Gallimore, for several useful 
suggestions.
This paper is based upon work supported by NASA grant NAG5-4826.  
JJF acknowledges support from R. Ciardullo through grant 
number GO-0612.01-94A from the Space Telescope Science Institute. 
ACF acknowledges the Royal Society for support, and 
KI acknowledges the PPARC for support. 

\pagebreak

\pagebreak


\begin{figure}
\epsscale{0.5}
\plotfiddle{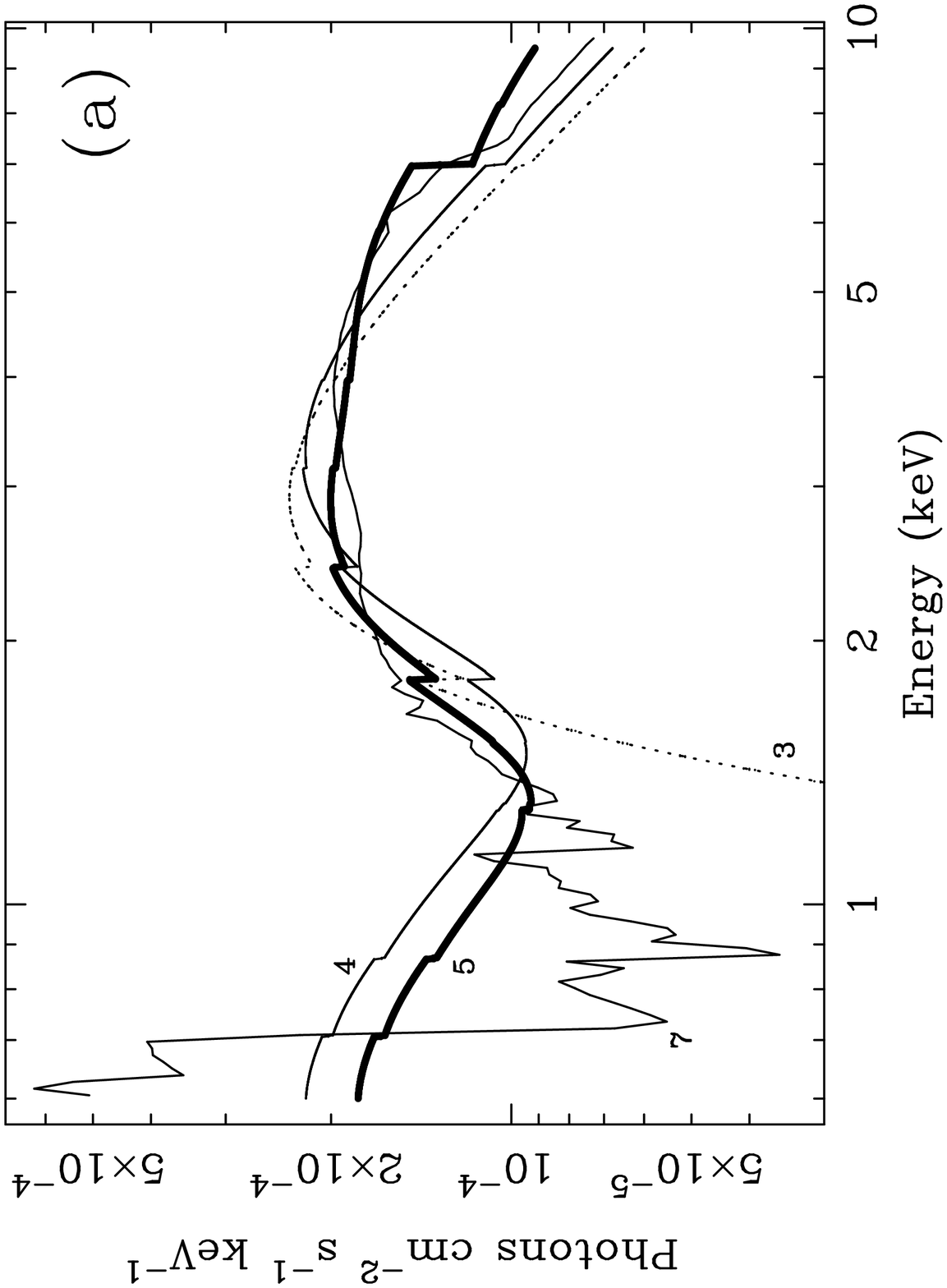}{100pt}{-90}{50}{50}{-200}{170}
\plotfiddle{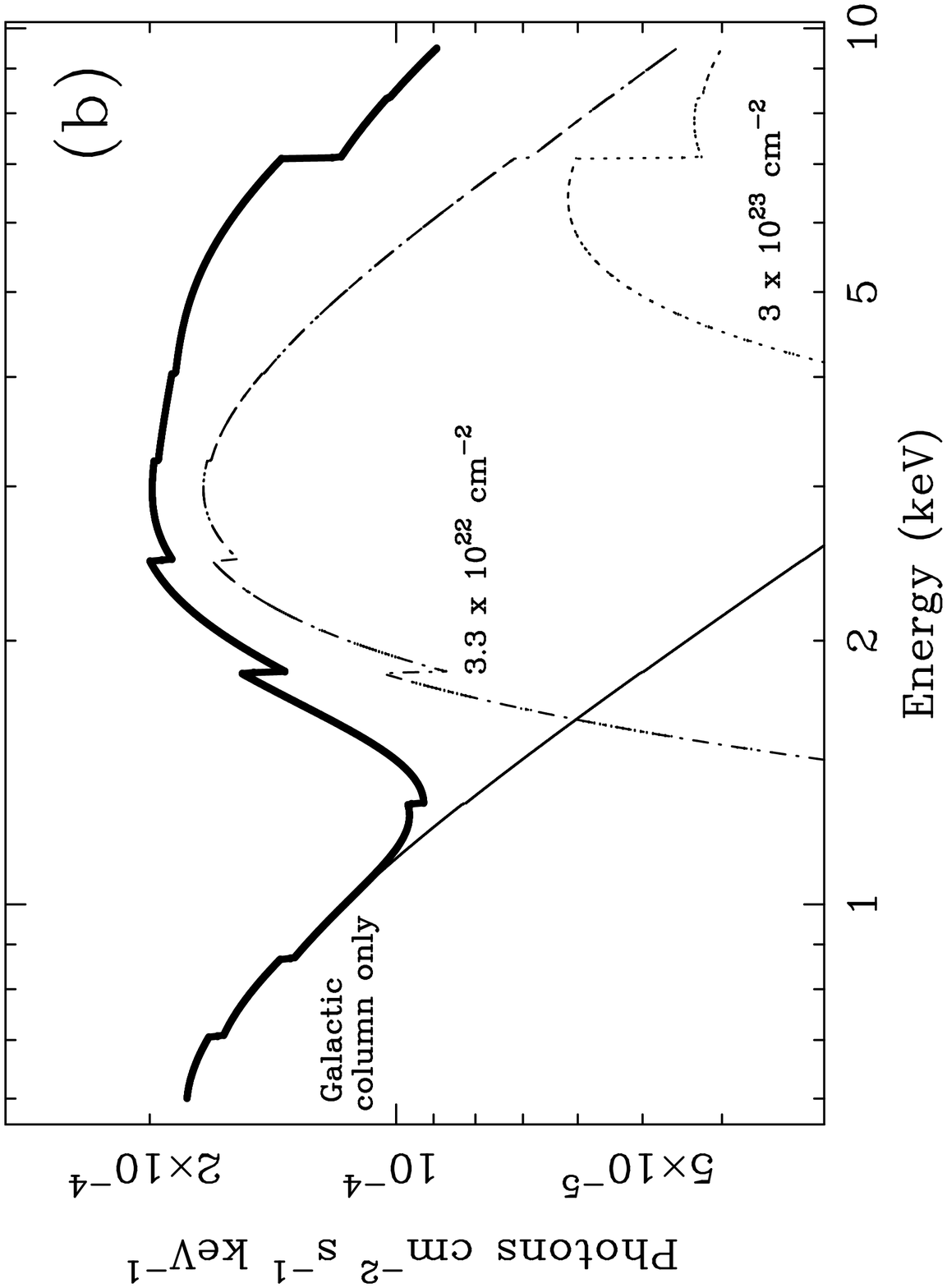}{100pt}{-90}{50}{50}{-200}{30}
\vspace{4.15in}
\caption{(a) Comparison of four of the spectral models fit to the \m6 data
(see the text for descriptions of these models). The numbers of the models
under comparison are given in the figure itself. The data are fit in the
0.6--9.5~keV band.
(b) Decomposition of Model~5 (shown as the thick solid line) into 
three individual absorbed power laws. 
Each power law represents one of the lines of sight discussed in \S 2.3.2. 
One power law is absorbed by only the Galactic column, 
one is absorbed by an intrinsic column of $\approx 3.3\times 10^{22}$~cm$^{-2}$ and one is absorbed by an intrinsic 
column of $\approx 3\times 10^{23}$~cm$^{-2}$. 
\label{fig1}}
\end{figure}
\pagebreak
  

\begin{figure}
\epsscale{0.5}
\plotfiddle{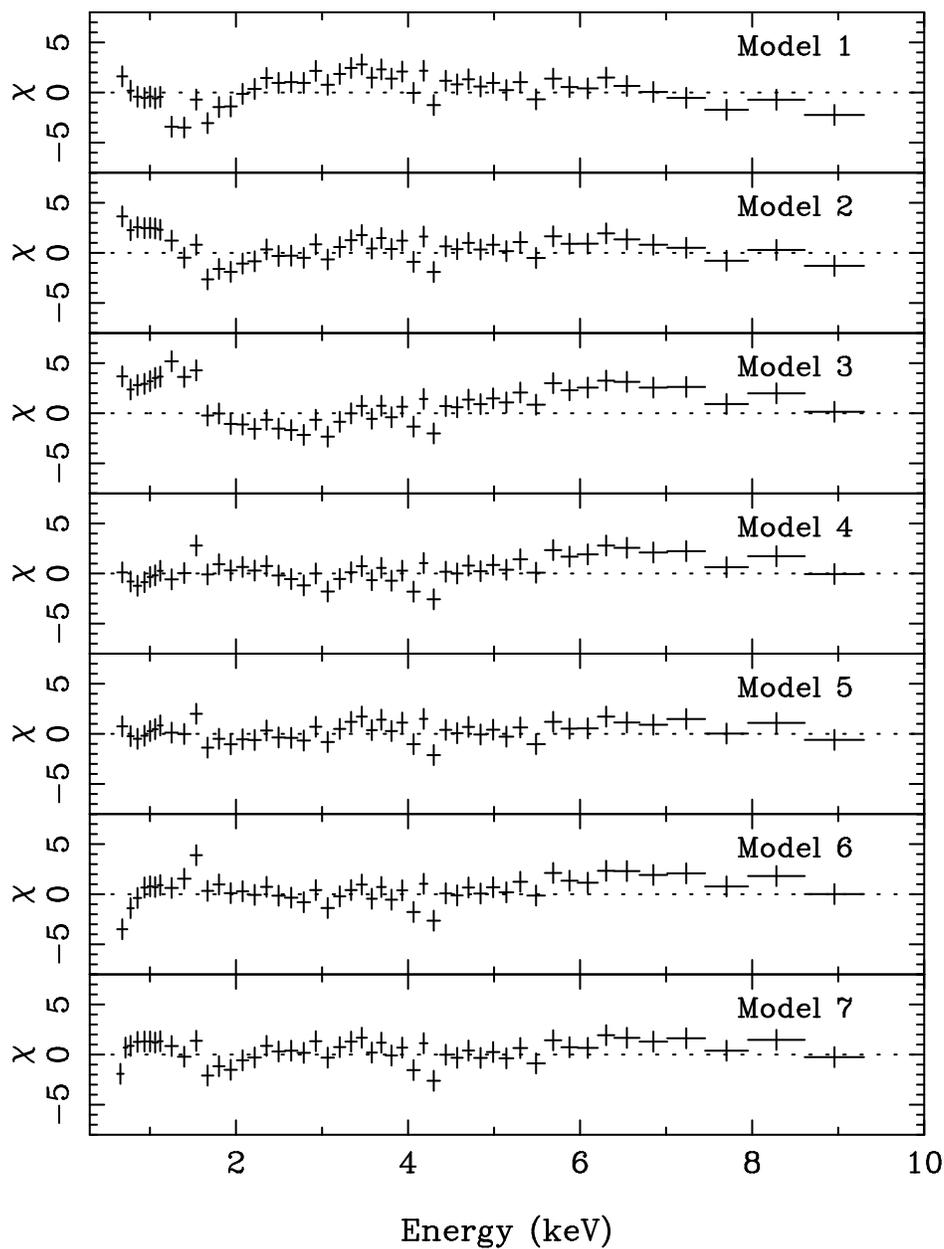}{340pt}{0}{70}{70}{-210}{-80}
\vspace{0.9in}
\caption{\ASCA SIS0 0.6--9.5~keV fit residuals for Models~1--7 (see the text for 
descriptions of these models). The ordinates for the panels (labeled $\chi$) show the 
fit residuals in terms of sigmas with error bars of size one. Only the SIS0 residuals are shown
for clarity, but the residuals for all the detectors are in good general agreement.
\label{fig2}}
\end{figure}
\pagebreak


\begin{figure}
\epsscale{0.5}
\plotfiddle{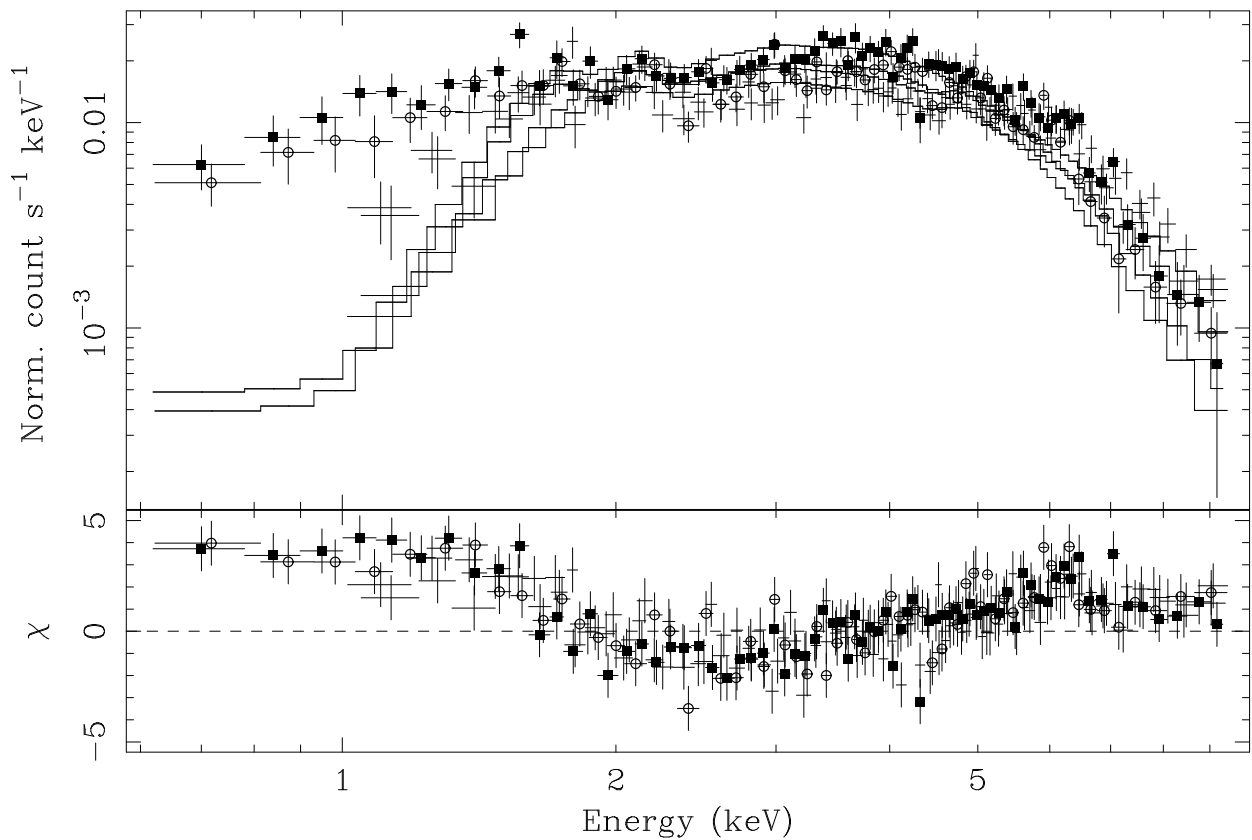}{250pt}{-90}{70}{70}{-270}{450}
\vspace{0.3in}
\caption{ \ASCA SIS0 (solid squares), SIS1 (open circles) and 
GIS (plain crosses) spectra of \m6 (shown in the observed frame). 
A model consisting of a power law, Galactic absorption,
and intrinsic absorption in \m6 has been fit to the data (Model 3).  
As described in the text, the power-law photon index is constrained to
be $\geq 1.6$. The ordinate for the lower panel (labeled $\chi$) shows the fit 
residuals in terms of sigmas with error bars of size one. Note the
clear systematic residuals throughout the spectrum.
\label{fig3}}
\end{figure}
\pagebreak


\begin{figure}
\epsscale{0.5}
\plotfiddle{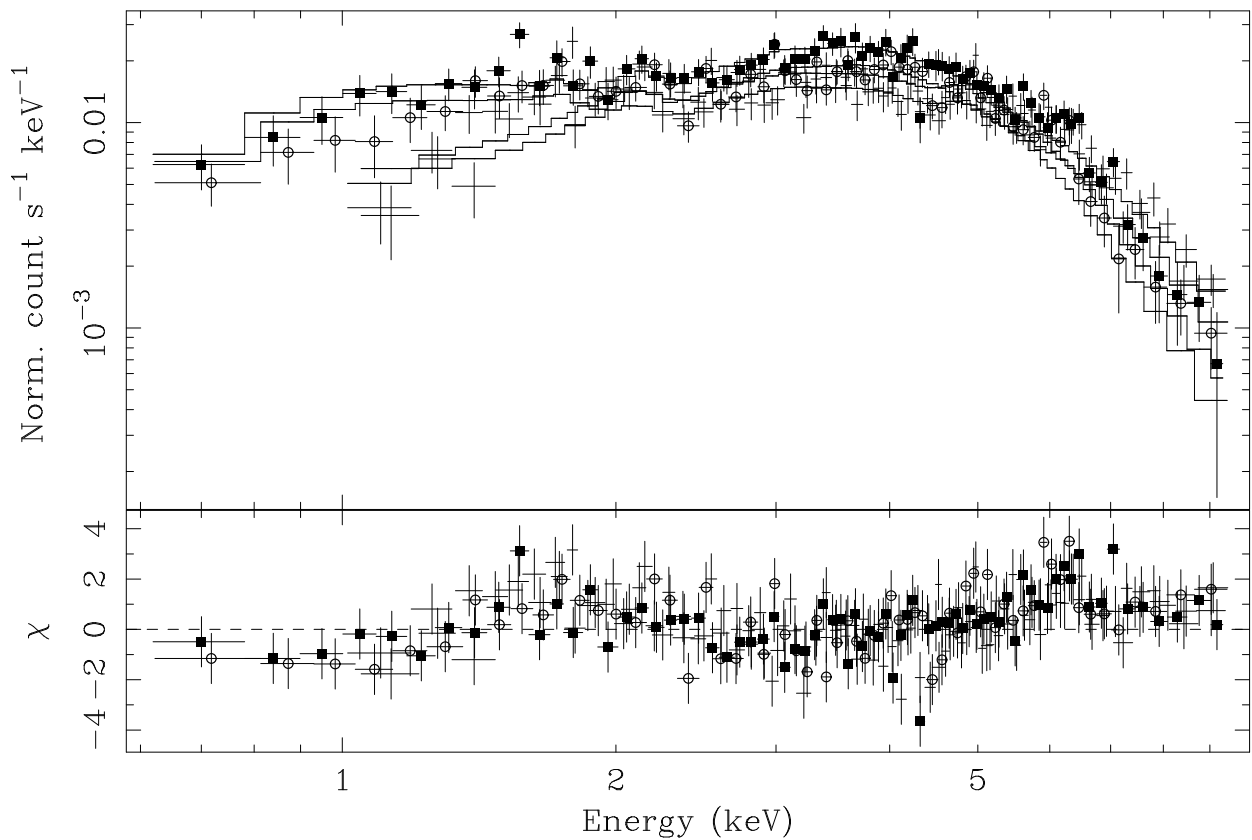}{250pt}{-90}{70}{70}{-270}{450}
\vspace{0.3in}
\caption{\ASCA SIS0 (solid squares), SIS1 (open circles) and 
GIS (plain crosses) spectra of \m6 (shown in the observed frame).
A model consisting of a power law, Galactic absorption and intrinsic 
partial covering has been fit to the data 
(Model 4).  The ordinate for the lower panel 
(labeled $\chi$) shows the fit residuals in terms of sigmas with 
error bars of size one. Note the `wave-like' systematic residuals in this
model. Also note the systematic positive residuals at the energy
of the iron~K$\alpha$ line. 
\label{fig4}}
\end{figure}
\pagebreak


\begin{figure}
\epsscale{1.0}
\plotone{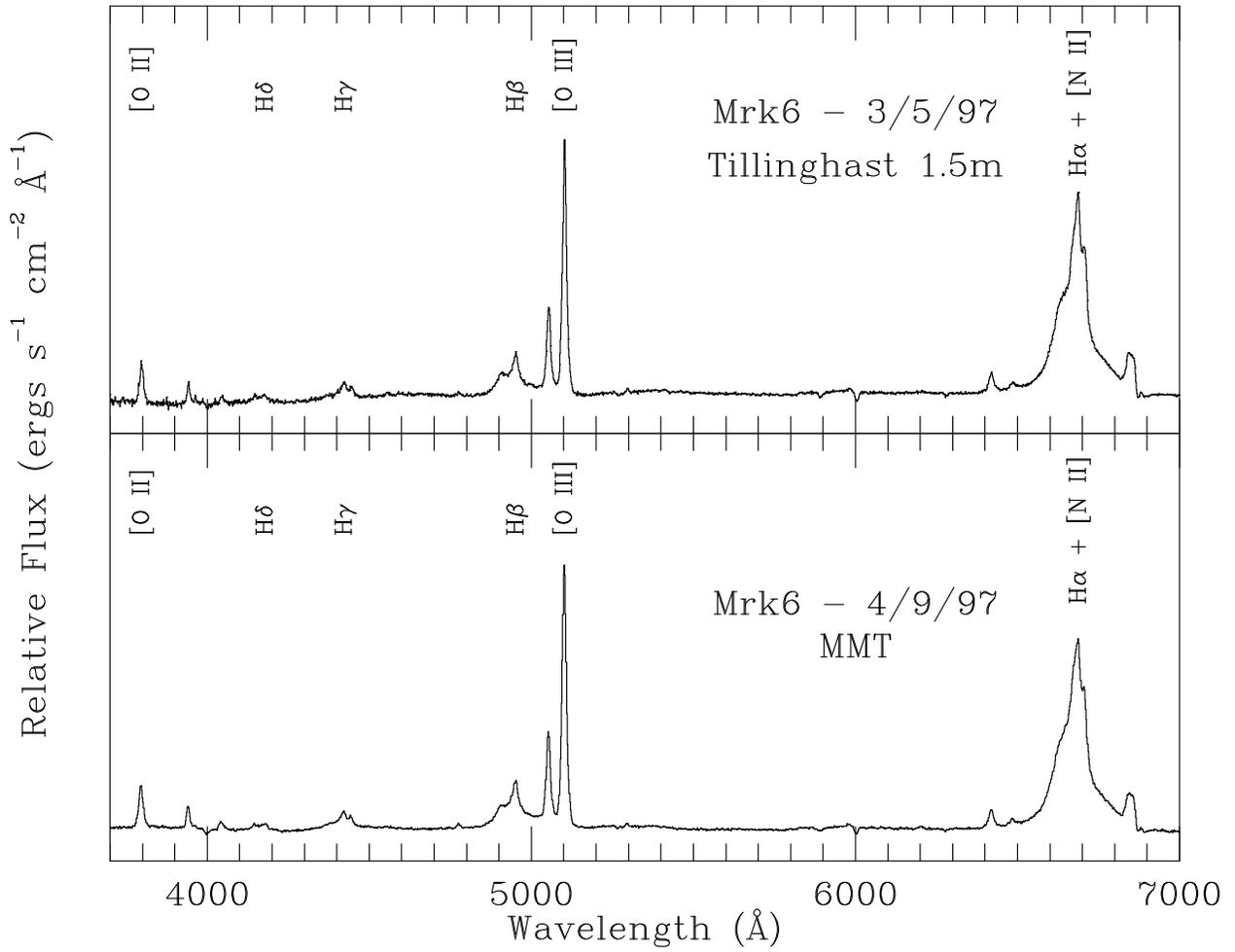}{}
\caption{Optical spectra of \m6 taken with the Tillinghast 
1.5~m and the MMT. The locations of prominent emission lines are shown.  
Note the broad blue-shifted components of H$\beta$ and H$\alpha$, 
though H$\alpha$ is also blended with [{\NIIf}].  
\label{fig5}}
\end{figure}
\pagebreak


\begin{figure}
\epsscale{0.5}
\plotfiddle{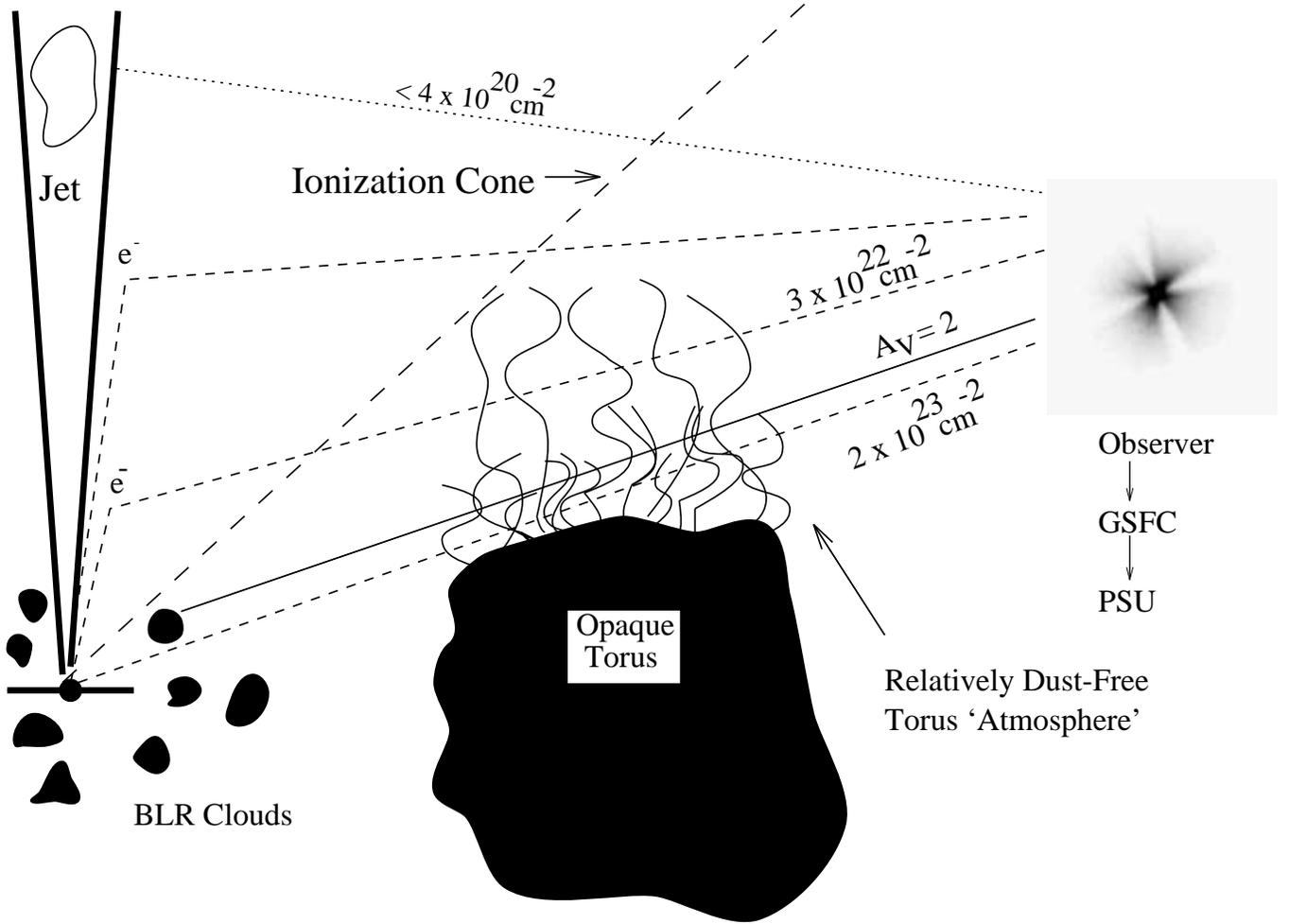}{300pt}{-90}{70}{70}{-270}{380}
\vspace{0.2in}
\caption{Sketch of the absorption geometry in \m6 for Model~9. 
The short-dashed lines show the X-ray lines of sight, 
the solid line shows the near-infrared/optical line of sight for one BLR cloud, and
the dotted line shows a radio line of sight. 
The near-infrared, optical and X-ray radiation intercepts
a relatively dust-free torus `atmosphere' that blocks ionizing radiation
but allows near-infrared/optical radiation to pass through with only
moderate extinction. This torus atmosphere is thereby responsible for
collimating the observed ionization cone (one edge of which is 
shown by the long-dashed line). 
We list the fitted X-ray absorption column densities next to the 
corresponding short-dashed lines (see Table~1), and we also 
show the $A_V$ from Rix et~al. (1990) next to the
near-infrared/optical line of sight. 
The strongest radio emission does not originate from the immediate vicinity
of the nucleus. 
The grayscale image at the right hand side shows the \ASCA SIS image. 
This sketch is not to scale. 
\label{fig6}}
\end{figure}
\pagebreak


\begin{deluxetable}{lccccc}
\tablenum{1}
\tablewidth{0pt}
\tablecolumns{6}
\tablecaption {\ASCA Spectral Fitting Results}
\small
\tablehead{ \multicolumn{6}{c}{Model Description and Number }  \\
\colhead{Parameter} & \colhead {PL + GA} & \colhead {PL + GA} &
\colhead{PL + GA} & \colhead {PL + GA} & \colhead {PL + GA}\\
\colhead{Name} & \colhead{ + ZA} & \colhead{ + PC} & 
\colhead { + 2 PC} &\colhead { + PC + WA}
& \colhead{ + 2 PC + Fe} \\ 
%
%
& \colhead{2} & \colhead{4} & \colhead{5} & \colhead{7} & 
\colhead{9}}
\startdata
\vspace{-4pt} Galactic & 5.4 & 7.4 & 7.4 & 7.4  & 7.4 \nl
$N_{H}$/($10^{20}$ cm$^{-2}$) & \nl
Photon Index, $\Gamma$ & $0.56^{+0.07}_{-0.09}$& 1.6 \tablenotemark{a} 
& 1.6 \tablenotemark{a} & 1.6 \tablenotemark{a} & 1.6 \tablenotemark{a} \nl
\vspace{-4pt}Redshifted & $1.19^{+0.16}_{-0.15}$ & --- & --- & --- 
& --- \nl
$N_{H}$/($10^{22}$ cm$^{-2}$)& \nl
\vspace{-4pt}Partial Covering (1) & --- & $4.90^{+0.29}_{-0.27}$ 
& $3.33^{+0.29}_{-0.16}$ &  $12.3^{+4.2}_{-2.7}$  
& $3.09^{+0.42}_{-0.79}$ \nl
$N_{H}$/($10^{22}$ cm$^{-2}$)& \nl
\vspace{-4pt}Partial Covering (1) & --- 
& $93.6^{+0.5}_{-0.5}$ &  $92.7^{+1.1}_{-0.7}$ &  $63.4^{+5.1}_{-4.4}$ 
&  $92.1^{+1.2}_{-2.1}$\nl 
Percentage & \nl
\vspace{-4pt}Partial Covering (2) & --- & --- & $29.8^{+4.4}_{-9.9}$ 
& --- &  $18.28^{+3.6}_{-8.7}$\nl
$N_{H}$/($10^{22}$ cm$^{-2}$) & \nl
\vspace{-4pt}Partial Covering (2) & --- & --- & $55.7^{+5.5}_{-2.9}$& 
--- & $50.2^{+9.2}_{-7.2}$\nl
Percentage \nl
$\xi$ (erg cm s$^{-1}$)  & ---  & ---  & --- & $51.9^{+5.7}_{-4.4}$     & --- \nl
$\log(N_{\rm H,warm}$)   & ---  & ---  & --- & $22.58^{+0.05}_{-0.07}$  & --- \nl
%
%
Line energy & --- & --- & --- & --- & 6.4 keV (fixed) \nl
Line sigma (eV) & --- & --- & --- & --- & $235^{+327}_{-168}$\nl
Line norm. ($\times 10^{-5})$ \tablenotemark{b} 
& --- & --- & --- & --- & $4.6^{+2.5}_{-2.5}$ \nl
$\chi^{2}$ / d.o.f. & 839.3 / 777 & 842.5 / 776 & 714.3 / 774  & 755.0 / 774  &
697.9 / 769 \nl
$\chi^{2}_{\nu}$ / P($\chi^{2}$ $\mid$ $\nu$) 
& 1.08 / 0.94 & 1.09 / 0.95 & 0.923 / 0.06 
& 0.975 / 0.32 & 0.908 / 0.03 \nl
\enddata
\tablenotetext{}{Key: PL = power law, GA = Galactic absorption (constrained
to lie within the radio-determined range; see the text), 
ZA = redshifted absorption, 
PC = partial covering, 
WA = warm absorber/emitter model (see the text for details), 
Fe = iron line.  All errors are given 
at the 90\% confidence level for one parameter of interest 
($\Delta \chi^{2} = 2.71$). P($\chi^{2}$ $\mid$ $\nu$) is the
chisquared rejection probability as defined in section~6.2
of Press et~al. (1989), and larger values of P($\chi^{2}$ $\mid$ $\nu$)
indicate statistically poorer models.}
\tablenotetext{a}{The photon index was constrained to lie within the range 
1.6--2.2.  However, in all models, the best-fitting value was found
when the photon index was 1.6 (see the text for further discussion).}
\tablenotetext{b}{The line normalization is quoted for the SIS0 detector.}
\end{deluxetable}

\begin{deluxetable}{lcccccl}
\tablenum{2}
\tablewidth{0pt}
\tablecaption {Details of Optical Observations}
\small
\tablehead{ 
\colhead{Observation} & \colhead {Telescope and} 
& \colhead{Range}  & \colhead{Dispersion \tablenotemark{b}} 
& \colhead{Aperture} & \colhead{Exposure} & \colhead{P.A.} \\
\colhead{Date} & \colhead{Spectrograph \tablenotemark{a}} & 
\colhead{(\AA)} & \colhead{(\AA)} & 
\colhead{} & \colhead{Time (s)}}

\startdata
5 March 1997 & Tillinghast 1.5~m --- FAST  & 3650--7500 & 2.95  
& $3\parcsec \times 180\parcsec$  & 900 & $90^\circ$\nl
9 April 1997 & MMT --- Blue Channel  & 3200--8630 
& 3.92 & $2\parcsec \times 150\parcsec$ 
& 300 & $-54^\circ$ \nl
\enddata

\tablenotetext{a}{The Tillinghast 1.5~m and Multiple Mirror Telescope (MMT) 
are located on Mount Hopkins in Arizona. The FAST spectrograph 
is described in Fabricant \etal (1998),
and the Blue Channel spectrograph is described in 
Schmidt, Weymann, \& Foltz (1989).} 
\tablenotetext{b}{Dispersion is \AA~per two pixel resolution element.}
\end{deluxetable}

\end{document}